\documentclass[conference]{IEEEtran}
\IEEEoverridecommandlockouts
\usepackage{cite}
\usepackage{amsmath,amssymb,amsfonts}
\usepackage{algorithmic}
\usepackage{url}
\usepackage{graphicx}
\usepackage{textcomp}
\usepackage{amssymb}
\usepackage{xcolor}
\usepackage{mdframed}
\usepackage{pifont}
\usepackage{enumitem} 
\usepackage{amssymb} 
\usepackage{array}
\usepackage{fancyhdr}
\usepackage{graphicx}
\usepackage{marvosym}
\usepackage{pgfplots}
\usepackage{lipsum}
\usepackage{tikz}
\usepackage{listings}
\usepackage{xcolor}
\usepackage[markup=default]{changes}
\usepackage{mathpazo}
\usepackage{lmodern}
\usepackage[T1]{fontenc}
\usepackage{caption}
\def\BibTeX{{\rm B\kern-.05em{\sc i\kern-.025em b}\kern-.08em
    T\kern-.1667em\lower.7ex\hbox{E}\kern-.125emX}}
\usepackage{tikz}
\usetikzlibrary{patterns}
\pgfplotsset{compat=1.18}
\begin{document}

\title{Combining GPT and Code-Based Similarity Checking for Effective Smart Contract Vulnerability Detection\\
\thanks{Identify applicable funding agency here. If none, delete this.}
}

\author{\IEEEauthorblockN{Jango Zhang}
\textit{Chongqing University}\\
202314021008@stu.cqu.edu.cn}

\maketitle
\begin{abstract}
\sloppy
With the rapid growth of blockchain technology, smart contracts are now crucial to Decentralized Finance (DeFi) applications. Effective vulnerability detection is vital for securing these contracts against hackers and enhancing the accuracy and efficiency of security audits. In this paper, we present SimilarGPT, a unique vulnerability identification tool for smart contract, which combines Generative Pre-trained Transformer (GPT) models with Code-based similarity checking methods.

The main concept of the SimilarGPT tool is to measure the similarity between the code under inspection and the secure code from third-party libraries. To identify potential vulnerabilities, we connect the semantic understanding capability of large language models (LLMs) with Code-based similarity checking techniques. We propose optimizing the detection sequence using topological ordering to enhance logical coherence and reduce false positives during detection. Through analysis of code reuse patterns in smart contracts, we compile and process extensive third-party library code to establish a comprehensive reference codebase. Then, we utilize LLM to conduct an in-depth analysis of similar codes to identify and explain potential vulnerabilities in the codes. The experimental findings indicate that SimilarGPT excels in detecting vulnerabilities in smart contracts, particularly in missed detections and minimizing false positives.


\end{abstract}


\section{Introduction}
Since the origin of Ethereum\cite{ethereum_main_page}, smart contracts have emerged as a fundamental element of blockchain technologies. Their immutable, transparent, and open-source characteristics have established the foundational Decentralized Finance (DeFi) application framework. Given that the DeFi ecology encompasses a multitude of cryptocurrencies \cite{defillama_totallocked}, it is imperative to identify and address security vulnerabilities in smart contracts. As reported by Defillama Hacks, hacking incidents have resulted in approximately \$9.06 billion in damages through November 2024\cite{defillama_hacks}. This scenario presents a substantial risk to the security of the entire DeFi ecosystem and user assets.Smart contract vulnerabilities primarily stem from design flaws and coding errors within decentralized applications, with logical inconsistencies representing a key attack vector for malicious hackers\cite{web3bugs}. Existing analysis tools\cite{mythril, slither, Confuzzius_paper,oyente} primarily focus on detecting vulnerabilities that follow predictable static patterns in control and data flows, such as reentrancy issues\cite{rodler2018sereum} and integer overflow vulnerabilities\cite{tan2022soltype}. However, these conventional program analysis approaches have shown limited effectiveness in practice\cite{chaliasos2024smart}.

The emergence of generative large language models \cite{gpt-4_report, A_survey_of_large_language_models} recently has revealed their significant advantages for auditing smart contracts. Recent investigations \cite{gptscan,gptlens,llm4vuln,propertygpt, Do_you_still_need} have demonstrated that large language models (LLMs) are promising in auditing smart contracts. Current research in LLM-based vulnerability detection primarily examines smart contracts at the contract or function level, using structured prompts as LLM inputs. The effectiveness of using LLMs for smart contract vulnerability identification is optimized through a two-phase approach that separates the initial detection process from the subsequent analysis of vulnerability root causes. However, this approach heavily relies on Large Language Models' inherent reasoning abilities and their pre-training knowledge base for detecting vulnerabilities in smart contracts \cite{llm4vuln}. This challenge can be mitigated by improving the model's reasoning skills,like fine-tuning. Yet, fine-tuning has limited scalability, risks overfitting to specific datasets, and cannot update knowledge in real-time.

Additionally, research by \cite{llm4vuln} revealed that simply incorporating vulnerability knowledge into LLMs is insufficient for improving their reasoning capabilities. The study demonstrated that LLMs can effectively reasoning for vulnerability detection when the vulnerability knowledge is carefully structured and systematically integrated.

Recent research \cite{111,112,code_reuse} has demonstrated that in the development of Defi, there is a considerable amount of code reuse due to developing costs, code security, and developer habits, and other reasons \cite{111}. This trend is particularly prominent in the development of Solidity smart contracts. The study \cite{code_reuse} analyzed over 350,000 Solidity smart contracts to investigate the present state of smart contract composition and code reuse. The study indicated that more than 80\% of subcontracts came from external sources, with Node Package Manager(NPM) as the largest external source, accounting for more than 72\% of external subcontracts and less than 17.06\% of subcontracts being self-developed. This conclusion emphasizes the dependency on third-party libraries and frameworks for smart contract development. This shows that smart contract development is significantly dependent on third-party package on npm.

Moreover, \cite{code_reuse} discovered that approximately 50\% of these self-developed subcontracts have fewer than 10\% unique functions. Code reuse, including self-developed subcontracts, is common at the function level. This scenario may represent that when confronted with complex functional requirements; developers prefer to reuse existing snippets of code rather than writing them from scratch to save development time and potential security concerns. The report further indicates that despite Solidity's built-in \textit{import} statement for subcontracts, a significant majority (over 56\%) of duplicated subcontracts are sourced from NPM package rather than direct imports, indicating a certain amount of uncertainty on dependency management\cite{code_reuse}.

Similarity checking for vulnerability detection have been developed in response to the extremely high rate of code reuse on Ethernet\cite{huang2021hunting,liu2018eclone,smartembed}. The technique involves analyzing code patterns and structures to identify potential security vulnerabilities, such as insecure mathematical operations and access control vulnerabilities, through comprehensive code comparison techniques, including data flow and control flow analysis, syntactic examination, and various pattern matching methodologies for smart contract verification\cite{smartembed}. Similarity checking for vulnerability detection in contracts face significant limitations in capturing comprehensive contract characteristics - including syntactic elements, semantic meaning, and functional behaviors. This inadequate representation often leads to substantial detection gaps and incorrect vulnerability identifications\cite{smartembed}.

To this end, we present SimilarGPT, the first tool that combines GPT and Code-based similarity checking(CBSC) to find coding flaws in smart contracts. Given the significant code reuse problem in Ethereum \cite{code_reuse}, we use a deep learning model to vectorize the "correct" code (i.e., generic third-party code with no vulnerabilities) and the code to be detected. By measuring the similarity, we can use the "correct" code from related third-party libraries as a reference and feed it into LLMs with the code to be detected for detailed semantic-based detection. This strategy is motivated by the belief that "in school, students with average grades and love to learn will tend to compare their answers with those of high-scores students after completing their after-school homework in order to find out the shortcomings of their solutions".
 
However, LLMs could generate the illusion that "although this function itself is not vulnerable, we believe that this function is vulnerable because other functions it calls may be vulnerable, which affects the security of this function". To overcome this issue, SimilarGPT uses a topology-based sequence of function calls, effectively mitigating the impact of the LLM illusion.

To obtain high-quality data for use as similar code,We refer to \cite{code_reuse} to collect samples and use our own data improvement methods to filter out negative examples. We collected third-party packages as a dataset, by using \cite{code_reuse} as reference.  After processing, this dataset consists of the top 150 most frequently used npm and GitHub package, with 35705 files and 357050 functions in total. In the following steps, we polished each function by removing comments, whitespace, indentation, and line breaks before filtering it using hash matching. After those processing procedures, we have 83,321 selected functions. This strategy not only increases data quality, but it also provides us with a dependable reference codebase, which serves as a solid foundation for future code similarity detection and vulnerability detection. 

\textbf{RoadMap}. the rest of this paper is organized as follows. We first introduce the related background in \ref{sec:background}, and then present the detailed design of SimilarGPT in \ref{sec:design}. Then, we show the experimental setup and results in \ref{sec:evaluation}. After that, we discuss the related work and limitations in \ref{sec:related_work} and \ref{sec:THREATS}, respectively. Finally, \ref{sec:conclusion} summarizes the paper.

\section{BACKGROUND}
\label{sec:background}
\subsection{Smart Contracts and Their Vulnerabilities}
Smart contracts enable decentralized finance (Defi) \cite{defi} in blockchain transactions, eliminating the need for middleman. According to DeFiLlama \cite{defillama_totallocked}, the total locked-in value on the three major blockchain platforms - Ether, Solana, and Tron - has reached \$74 billion by November 2024. Smart contract vulnerabilities can lead to property loss, as seen in the \textit{TheDao} event, which cost around \$150 million. However, owing to the characteristic of the blockchain, The contract cannot be modified once deployed on the blockchain, leaving any vulnerabilities open to potential exploitation.\cite {zhou2023sok}.

\subsection{LLM-driven Formal Vulnerability Detection of Smart Contracts}
Generative pre-trained transformer (GPT) models, such as GPT-4 \cite{gpt-4_report}, are extensive language models (LLMs) developed using vast datasets. These models possess the ability to learn from text, comprehend and analyze source code, and perform zero-shot learning \cite{zero_shot}, enabling them to detect security vulnerabilities in code without needing specific examples.

Despite its promise for code auditing, multiple studies have demonstrated that GPT models cannot replace human auditors. David et al. (2023) shows that inputting the entire project code into a GPT model is both expensive and difficult for achieving accurate detection outcomes. Sun et al. \cite{ sun2024llm4vuln} studied the effect of different components (e.g., function calls, external knowledge) on vulnerability detection for LLM. Even under optimal conditions, such as improving GPT-4's vulnerability knowledge with Retrieval Augmentation Generation (RAG), the accuracy in the context of LLM's vulnerability detection paradigm remains below 30\% when both the decision (i.e., correctly determining the presence of a vulnerability in the code) and the argument (i.e., correctly pointing out the type of vulnerability) are correct. 

\subsection{similarity-based code detection}

Code-based similarity checking (CBSC) techniques identify potential vulnerabilities by analyzing and comparing the specific code implementations of different smart contracts. The core idea of this technique is to perform code pattern matching for similar structures using various methods, such as data flow/control flow analysis\cite{huang2021hunting}, semantics analysis\cite{smartembed}, and symbolic execution\cite{liu2018eclone}. The goal is to identify the potential presence of similar code fragments that may contain known vulnerabilities \cite{smartembed}. For instance, certain code patterns (e.g., insecure mathematical operations and access control vulnerabilities) may be repeated in multiple contracts. 

Nevertheless, it is challenging to capture the syntactic, semantic, and functional information of contracts through the abstracted characterization of contracts using CBSC techniques. Therefore, it is likely to result in many false positives and underreporting \cite{smartembed}.

\section{DESIGN OF SIMILARGPT}
\label{sec:design}
\begin{figure*}[ht]
    \centering
    \includegraphics[width=\textwidth]{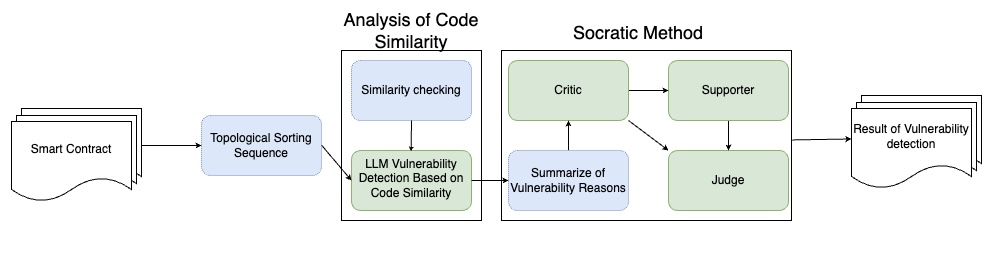}
    \caption{\textbf{An overview of SimilarGPT, green blocks indicating GPT works and green blocks suggesting code similar analysis.}}
    \label{fig:main_fig}
\end{figure*}
The overall design of SimilarGPT and the primary challenges of LLM in detecting smart contract vulnerabilities are depicted in \ref{AA}. The three critical components of SimilarGPT are then introduced in \ref{BB}, \ref{CC}, and \ref{DD}, respectively.

\subsection{overview and challenges}\label{AA}
The entire framework of SimilarGPT is illustrated in Fig. \ref{fig:main_fig}. The green box represents the LLM-based agent, while the blue box represents the data processing. The smart contract code is first preprocessed by breaking it into distinct functions, ensuring each data segment to be evaluated is a separate function. The detection order of these functions is established through topological ordering based on their calling relationship. Subsequently, the detector is provided with similar code from third-party package to conduct a similarity checking for the code. The Socrates section receives the detection result for subsequent evaluation.

\textbf{Challenge}. Despite the concise architecture of SimilarGPT in Fig. \ref{fig:main_fig}, it is challenging to conduct effective smart contract auditing, which involves enhancing the detection rate while minimizing the impact of LLMs' illusions. Several challenges were encountered during the design and implementation of SimilarGPT, as detailed below:
\begin{itemize} 
\item \textit{How can SimilarGPT effectively detect smart contract vulnerabilities?} While some research has demonstrated the potential of LLMs for smart contract vulnerability detection, most existing tools cannot combat the increasingly difficult hacking event on the blockchain due to the lack of well-designed knowledge and the rapid iteration of new smart contract vulnerability . This work proposes a GPT-driven Code-based similarity checking for vulnerability detection for smart contract. We will first present this in \ref{BB}.

\item \textit{How can SimilarGPT effectively handle contextual information?} The previous paper \cite{llm4vuln} explored how various factors affect LLMs' ability to detect vulnerabilities in smart contracts, including the impact of model hyperparameter configurations \cite{_hyperparameters} and the integration of external knowledge sources \cite{_rag}. According to \cite{llm4vuln},simply providing context info may not always help LLMs' reasoning about vulnerabilities. It may also cause diversions, preventing LLMs from correctly discovering vulnerabilities. So, how to effectively convey vulnerability-related context will significantly LLMs' capacity to detect vulnerabilities. We combine a sequence of vulnerability detection functions based on topological ordering into SimilarGPT. It is used to deal with the large model's contextual information processing during the smart contract detection procedure. This will be given in \ref{CC}.

\item \textit{How to collect a high quality dataset for code comparison?} Acquiring a high-quality dataset remains a critical foundation, both for conventional vulnerability detection methods that rely on code similarity and for our SimilarGPT approach. We refer to \cite{code_reuse}'s method to obtain the dataset and use our own data filtering method to find out the possible negative samples in the dataset. This will be \ref{DD}

\item \textit{How Can We Reduce False Positive Rate in Large Model Vulnerability Detection?} False positives rate play a critical role in vulnerability detection. However, the hallucination of LLMs \cite{llm_hallucination} constantly leads to false positives for vulnerabilities. As a result, understanding how to reduce the impact of the LLMs illusion becomes critical. We will use recent developments in the Multi-agent Framework in LLMs  to connect this topic. Specific details will be presented in \ref{EE}. 

\end{itemize}

\subsection{Code-based similarity checking}\label{BB}
\captionsetup{justification=raggedright,singlelinecheck=false} 
\begin{figure}[ht]
    \centering
    \includegraphics[width=1\linewidth]{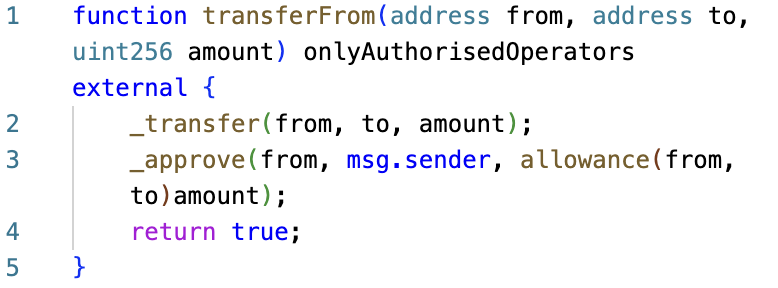}
    \caption{The Redacted Cartel exploit}
    \label{fig:Redacted_vul}
\end{figure}
\begin{figure}[ht]
    \centering
    \includegraphics[width=1\linewidth]{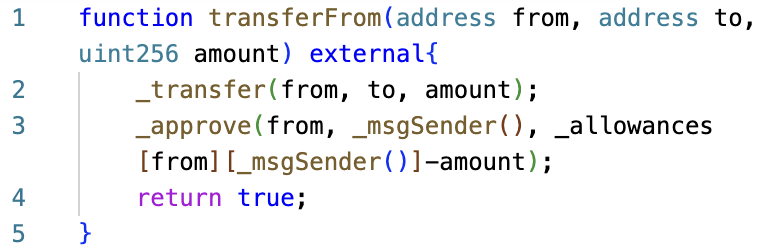}
    \caption{transferFrom function in Openzeppelin's ERC20 contract}
    \label{fig:ERC20}
\end{figure}

Fig. \ref{fig:Redacted_vul} shows the first code sample from \cite{web3bugs}, which focuses on vulnerabilities that machines cannot audit. \cite{web3bugs} demonstrates that in Fig. \ref{fig:Redacted_vul}, The problem described is difficult to identify, having passed through several rounds of manual auditing and tool evaluation without being discovered. This is because it needs to know the meaning of \_allowances, the purpose of the transferFrom function, and the business model. Fortunately, an ethical hacker discovered the vulnerability and reported it to Immunefi \cite{immunefi}, a web3 vulnerability bounty site. The project side of Redacted Cartel rewarded the hacker with around \$560,000. 

However, the code was probably a clone from the Openzeppelin package. The \textit{transferFrom} function for openzeppelin's older versions of ERC20 is shown in Fig. \ref{fig:ERC20}, and the two functions are largely similar, differing primarily in their modifiers\footnote{Similar to \cite{web3bugs}, we simplify the actual code to clarify it.}.  However, the vulnerability can be discovered easily by means of similarity checking. However, simply using CBSC for vulnerability detection makes it hard to capture contracts' syntactic, semantic, and functional information. 

Inspired by this, We use the similar correct code as a reference and use GPT attempt to identify potential vulnerabilities by comparing the differences between the correct third-party package's code and the code to be identified. The difficulty of understanding function semantics of traditional CBSC is overcome by GPT.

\subsection{Topological Ordering-based Vulnerability Detection Sequencing}\label{CC}

\begin{figure}[ht]
    \centering
    \includegraphics[width=0.8\linewidth]{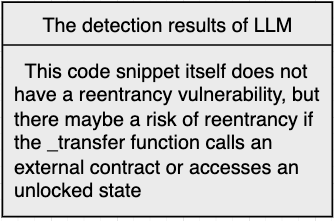}
    \caption{Detection results of the transferFrom function in Fig. \ref{fig:ERC20}}
    \label{fig:huan}
\end{figure}

In this paper,Fig. \ref{fig:huan} depicts the outcome of using the code in Fig. \ref{fig:ERC20} as a direct input to GPT-4 for vulnerability detection. When conducting vulnerability assessments of Large Language Models (LLMs), we frequently encounter scenarios similar to those illustrated in Fig. \ref{fig:huan}. The function in Fig. \ref{fig:ERC20} has been audited by some specialists. Thus, we can presume that it has no vulnerability. Similarly, no apparent vulnerabilities were found when LLM examined the transferFrom function. However, it indicated that other functions called by the transferFrom function, such as the function \_transfer in Fig. \ref{fig:ERC20}, are vulnerable. In the end it was determined that this case was considered vulnerable. This is particularly prevalent in the detection of reentrant vulnerabilities. Predictably, there are likely two causes for this predicament. 1. Insufficient context information. 2. Incoherent reasoning resulting from the hallucination of LLMs. We focus on solving the first problem.

It is evident that if LLM thinks that function B, called by function A, has a vulnerability, we can test function B first. Then, we can test function B first and then function A. In the same way,  if function B calls function C, then we test function C first. Then, there will always be a function at the bottom of the list which does not call any other function. Following this, a directed acyclic graph of function calls exists in function calls. Function A is called function B, and A depends on B. So, at the time of detection, B was detected first. In this way, we can perform function-based smart contract detection in the sequence of topological ordering \cite{topo}. Of course, some tools can also be used \cite{surya} for call graph construction.

Let $G = (V, E)$ be a directed acyclic graph where $V$ represents the set of functions and modifiers in a smart contract, and $E$ represents the calling relationships between them. For any two vertices $v_i, v_j \in V$, an edge $(v_i, v_j) \in E$ indicates that function $v_i$ calls function $v_j$.

     Given that vulnerabilities in called functions can affect their callers, we propose a systematic detection approach based on topological ordering. Let $f: V \rightarrow V$ be a calling relationship where $f(v_i) = v_j$ denotes that $v_i$ calls $v_j$. Then:
      
     \begin{itemize}
         \item Let $S = \{v_1, v_2, ..., v_n\}$ be the set of all functions and modifiers in the contract.
         \item For each vertex $v_i \in S$, we define:
            \begin{enumerate}
                \item $C(v_i) = \{v_j \in S \mid f(v_i) = v_j\}$ as the set of functions called by $v_i$
                \item For each $v_j \in C(v_i)$, we add edge $(v_j, v_i)$ to $E$
            \end{enumerate}
         \item Let $\tau: V \rightarrow \mathbb{N}$ be a topological ordering of $G$ such that for every directed edge $(v_i, v_j) \in E$, $\tau(v_i) < \tau(v_j)$
     \end{itemize}
     
     This topological ordering $\tau$ provides the optimal sequence for vulnerability detection, ensuring that called functions are always analyzed before their callers. For any $v_i, v_j \in V$, if $v_i$ calls $v_j$, then $\tau(v_j) < \tau(v_i)$, guaranteeing a bottom-up analysis approach.

   As for some possible cases, such as rings in the sequence of function calls, we plan to optimize those scenarios in future work.

\subsection{High-quality Data Collection and Filtering}\label{DD}
\textbf{Collecting code from third-party libraries}. In order to employ similarity code for checking, obtaining an extensive amount of code from third-party packages is necessary. We have collected the most frequently used npm and GitHub packages in recent years\cite{code_reuse,111,112}. For our data collection methodology, we refer to research from \cite{code_reuse}, which analyzed over 350,000 smart contract source codes gathered from Etherscan, encompassing both Ethereum mainnet and Goerli testnet deployments between January 2021 and January 2023. A catalog of third-party packages was subsequently generated from the \textit{import} statements of the collated contracts. The 150 most frequently used third-party libraries will be our third-party library code for reference. The utilization of each third-party library between January 2021 and January 2023 was determined through \textit{import} statements and clone detection methods.

For these third-party library codes, we implemented the following approach. Collect the various versions of third-party libraries; for example, as of November 2024, the npm package of "@openzeppelin/contracts"\cite{openzeppelin} has 87 distinct versions, which equate to 87 gzip package files. For each third-party library we collect all the different versions of its gzip package. After collecting 150 distinct versions of third-party libraries, we gathered around \textbf{10,000} gzip package files. We collect all of the .\textit{sol} files in those gzip packages. Then, we have gathered around \textbf{46918} \textit{.sol} files. Moreover, we extract each function in the file and calculate its hash; this allows us to filter out most of the identical code, particularly for multiple versions of the same third-party library with just minor code differences. This way, the vulnerable code that has been modified will be stored. There are \textbf{766,505} functions before the hash match and only \textbf{35,709} functions following the hash-based filtering.

After hash matching, we use the \textit{all-MiniLM-L6-v2}\cite{huggingface_all-MiniLM-L6-v2} model based on \textit{sentence-transformers}\cite{sentence_transformer} to convert the code into vectors. all-MiniLM-L6-v2 model maps sentences and paragraphs to a 384-dimensional dense vector space and can be used for tasks like clustering or semantic search. The model is based on pre-trained nreimers/MiniLM-L6-H384-uncased model and fine-tuned in on a 1B sentence pairs dataset. The all-MiniLM-L6-v2 model is one of the best models based on Sentence Transformers\cite{sbert_all-MiniLM-L6-v2}. Compared to other models, \cite{sbert_1,sbert_2,sbert_3}  exhibits higher efficiency, and the encoding pace is also fast.          

\textbf{Similarity calculation}. Refer to \cite{smartembed}, we compute the Euclidean distance to assess code similarity. Specifically, we define the semantic distance and similarity between two code fragments $C_1$ and $C_2$, as well as their related code embeddings $e_1$ and $e_2$, as:
\[
\text{Distance}(C_1, C_2) = \frac{\text{Euclidean}(e_1, e_2)}{\|e_1\| + \|e_2\|}
\]

\noindent{Similarity is defined as follows:}
\[
\text{Similarity}(C_1, C_2) = 1 - \text{Distance}(C_1, C_2)
\]

Two code snippets $C_i$ and $C_j$ are considered to be clones if their similarity scores above a certain similarity threshold $\delta$. Based on our observations, 0.65 can be utilized as a similarity threshold. Above 0.65, we may consider the two codes just similar; below 0.65, the two codes maybe different in semantically, structurally, or functionally distinct.

In this method, we can group the code into three categories: code with similarity 1, code with similarity less than 1 but greater than 0.65, and code with similarity less than 0.65. 
\begin{enumerate}[label=\arabic*)]
    \item \textit{similarity = 1}.As expected, the code with a similarity of 1 is cloned by the developer from third-party libraries such as openzeppelin\cite{openzeppelin}. we will directly determine if there is a vulnerability in the code. 
    \item \textit{0.65 < similarity < 1}. For codes with similarity less than 1 and more than 0.65, which is the main focus of our research, we incorporate those similar codes into the GPT as part of the prompt. 
    \item \textit{similarity < 0.65}. For codes with a similarity of less than 0.65, no knowledge augmentation is performed, and they are inserted into the prompt as input to the GPT.
  \end{enumerate}

\textbf{Filter vulnerability code}. In the case that these third-party libraries contain potential vulnerability code. We employ the current vulnerability datasets for smart contract that have been collected from frequently used public datasets, including defihacks\cite{defihack}, slowmist\cite{slowmist}, and github's issues\cite{github}, according to \cite{llm4vuln}. For the current \textit{top 150 third-party libraries}, we attempt to gather the historical vulnerabilities that have occurred in those third-party libraries. Subsequently, we refer to the GPT-4-based method outlined in \cite{llm4vuln} to extract relevant vulnerability knowledge from the vulnerability descriptions. We attempt to annotate the corresponding original code by labeling it as a function with a vulnerability after carefully verifying the vulnerability's existence. Nevertheless, we do not abandon such vulnerable functions; they are employed as illustrations of vulnerabilities for subsequent testing. This is precisely the function of traditional code similarity-based vulnerability detection.

\subsection{Socratic method}\label{EE}
Previous research \cite{llm4vuln,gptlens,Do_you_still_need} discovered that vulnerability detection based on LLMs generates a substantial number of false positives, especially when the context is not given correctly \cite{llm4vuln}. Several issues contribute to these false positives, including knowledge mismatch and poor LLM reasoning ability. We are primarily focus with decreasing the false positives caused by LLMs illusions.

As LLM research advances, there is an increasing recognition of Prompt Engineering's ability to dispel LLMs' misconceptions. The Socratic method of debate\cite{soc_average,soc_Markov,soc_solo,soc_others} is very attractive. Asking and responding to questions, starting with generalized beliefs and then testing their internal consistency through rebuttals\cite{soc_solo}, helps us get closer to the truth, reducing the illusions that may be present in large-scale models and thus improving the accuracy of the models' judgments. In particular, we established three LLM roles: Critic, Supporter, and Judge \cite{soc_solo}. The following is the flow of their interactions:
\begin{itemize}
    \item \textbf{Detector} is a critical component to performing effective smart contract auditing. We feed the code to be detected and the similar code into Detector, and the vulnerabilities are identified by recognizing the differences in implementation between the two pieces of code.
    \item \textbf{Critic} \& \textbf{Supporter}. In order to determine the validity of the vulnerability reason during the detecting phase, we use the Socratic method to assess the validity of the output from the detecting phase. Specifically, Critic refutes the cause of the vulnerability given by Detector, while Supporter further evaluates Critic's output to debate and determine the most appropriate cause of the vulnerability.
    \item \textbf{Judge} is responsible for synthesizing the vulnerability cause, the vulnerability cause arguments given by Critic and Supporter. And, it independently gives its own judgment.
\end{itemize} 

In our future work we consider using multiple rounds of iterations to optimize the Socratic method.\cite{soc_Markov}\cite{soc_Markov}

\section{Evaluation}
\label{sec:evaluation}
\subsection{Experimental Setup}
\label{subsection:setup}
In this section, we will provide a summary of some key implementation details about SimilarGPT

\textbf{GPT model setups}. We use the GPT-4-turbo model from OpenAI for SimilarGPT. Our work indicates that GPT-4-turbo could be cost-effective while offering enough inference capacity. The model hyper-parameters are maintained at their default values, with TopP set to 1, presence penalty set to 0, and frequency penalty set to 0. Nonetheless, It should be mentioned that to lessen the impact of the GPT's output's unpredictability, we set the temperatures of the three roles-Critic, Supporter, and Judge—to 0. To retain some creativity, the detector's temperature is set to 0.8, which gives LLM several possibilities for produce some creative results\cite{llm4vuln}. Furthermore, each agent interaction occurs in a fresh session, ensuring independence between conversations and preventing any potential interference from previous responses.

\textbf{Datasets}. To accurately assess the detection capability of SimilarGPT and the efficacy of CBSC. We employ real-world vulnerabilities and smart contract audit reports from reputable industry companies as datasets. We collect two datasets separately. The first dataset comprises vulnerability data from \textit{Defihack}\cite{defihack}, a well-known DeFi Hacks dataset, and \textit{CVE} (Common Vulnerabilities and Exposures)\cite{cve}. We collect 13 typical vulnerability data from the Defihack and CVE vulnerability collections. We have included a variety of types of vulnerabilities, such as overflow, access control, bad randomness, price manipulation, and logic vulnerability, to ensure that the vulnerability data is comprehensive. Those vulnerabilities encompass a range of ages, from 18 to 24. The second dataset is sourced from \textit{Solodit}\cite{solodit}. This popular smart contract auditing website is specifically designed to facilitate the auditing of Web3 projects, with a particular emphasis on smart contract security. SimilarGPT will be assessed by collecting 67 vulnerable functions from \textit{Solodit} as positive samples and 71 negative samples (i.e., non-vulnerable code). The specific gathering methods are similar to those employed to address vulnerable code from third-party libraries.

\textbf{Research question}. Our proposed method focuses on two primary issues: enhancing recall and lowering false positive rate. Thus, we developed a series of experiments to  address the following research questions(RQs):
\begin{enumerate}
\item \textbf{RQ1:} How effective is SimilarGPT at detecting vulnerabilities? How does it compare to traditional vulnerability detection methods, including those based on LLM? 
\item \textbf{RQ2:} How helpful is the Socratic method for improving SimilarGPT's precision rate?
\item \textbf{RQ3:} How effective is SimilarGPT's Code-based similarity checking in improving the recall of SimilarGPT?
\end{enumerate}

\subsection{RQ1 - Vulnerability Detection}
We examined SimilarGPT's detection of \textit{real-world vulnerabilities}. These \textit{real vulnerabilities} once led to over \$1,000,000  losses in defi. The settings for SimilarGPT are essentially kept as specified in \ref{subsection:setup}, i.e., the parameters are kept as defaults, with only the detector role set to a temperature of 0.8 and the Critic, Supporter and Judge roles set to a temperature of 0. Table \ref{table:data} displays the performance of SimilarGPT on 13 \textit{real-world vulnerabilities}. SimilarGPT is compared to \textit{Slither}\cite{slither}, a popular static analysis tool for detecting vulnerabilities, and \textit{Mythril}\cite{mythril}, a symbolic execution tool. In addition, \textit{Gptlens}\cite{gptlens} is an adversarial framework for vulnerability detection in smart contracts based on the Large Language Model (LLM), aiming to overcome the accuracy and recall concerns of existing LLM tools in vulnerability detection.

\begin{table}[ht]
\centering
\resizebox{\columnwidth}{!}{
\begin{tabular}{c c c c c c} 
 \hline
 Contracts & Description & \rotatebox{70}{SimilarGPT} & \rotatebox{70}{GptLens} & \rotatebox{70}{Slither} & \rotatebox{70}{Mythril} \\ [0.5ex] 
 \hline
Ragnarok & access control & \ding{51} & \ding{51} & \ding{55} & \ding{55} \\ 
Nimbus Platform	& miscalculation & \ding{51} & \ding{55} & \ding{55} & \ding{55} \\ 
LuckeyTiger & bad randomness & \ding{51} & \ding{51} & \ding{51} & \ding{55} \\ 
ShadowFi	& access control & \ding{51} & \ding{55} & \ding{55} & \ding{55} \\ 
Grim Finance	& reentrancy & \ding{55} & \ding{55} & \ding{55} & \ding{55} \\ 
Bad Guys by RPF	& logic error & \ding{55} & \ding{55} & \ding{55} & \ding{55} \\
Uerii & access control & \ding{51}  & \ding{51} & \ding{55} & \ding{55}  \\  
GPU	& logic error & \ding{55} & \ding{55} & \ding{55} & \ding{55} \\ 
ZongZi	& price manipulation & \ding{55} & \ding{55} & \ding{55} & \ding{55} \\ 
Hopelend	& overflow & \ding{51} & \ding{55} & \ding{55} & \ding{55} \\ 
Coinlancer & access control & \ding{51} & \ding{51} & \ding{55} & \ding{55} \\ 
Virgo\_ZodiacToken & logic error & \ding{51} & \ding{55} & \ding{55} & \ding{55} \\ 
Lancer	& overflow & \ding{55} & \ding{55} & \ding{55} & \ding{51} \\ 
\hline
\end{tabular} 
} 
\caption{Vulnerability detection results for 13 real-world vulnerabilities.}
\label{table:data}
\end{table}

In Table \ref{table:data}, the first column lists the vulnerability name, followed by the corresponding vulnerability description. The remaining columns provide the detection results for each tool. Table \ref{table:data} shows that SimilarGPT outperforms all the tools, with SimilarGPT detecting 8 out of 13 vulnerabilities, followed by \textit{Gptlens} with 4 vulnerabilities, \textit{Slither} detecting only one vulnerability related to bad-randomness, and \textit{Mythril} detecting one vulnerability related to overflow. In those undetected vulnerabilities, the \textit{Lancer} and \textit{ZongZi} contracts were due to that they didn't provide enough contextual information, including the solidity version and calling function information. The vulnerabilities in the \textit{Bad Guys by RPF} contract were identified due to the absence of modifiers applied to its parameters. Yet Critic and Judge continue to argue that there is no vulnerabilities. The \textit{GPU} weakness was not identified as a perceived risk, and our think that this may be a result of the knowledge update time of large models. The knowledge of GPT-4-turbo is updated until 2023.

And the capabilities of SimilarGPT can be improved in real time by adding third-party package updates and open-source vulnerability code from different auditing platforms.

\textbf{Defi-fork-bugs}. The vulnerability found in the \textit{Uranium}\cite{uranium} contract led to losses of about \$50 million. This issue stemmed from a slightly modified function originally from \textit{uniswap v2}. A coding error introduced the bug, causing the significant financial loss.In fact, we can avoid this bug entirely. As early as April 2021, a identical vulnerability had impacted \textit{Nimbus}\cite{Nimbus}. If we're serious about learning this lesson. But even in the 2023,\textit{Swapos}\cite{swapos} contract experienced the same coding error, although the identical vulnerability had already occurred two years ago. In this time, \textit{Swapos}'s vulnerability resulted in losses of almost \$468,000. Nevertheless, this incident was completely avoidable, and the security weakness that led to it should never have existed in the first place. And, this is where SimilarGPT's effectiveness may manifest.

\begin{mdframed}[backgroundcolor=gray!30, hidealllines=true] 
    \textbf{Answer for RQ1:}SimilarGPT excels in detecting \textit{real-world vulnerabilities}, identifying a broader range of vulnerabilities more effectively compared to other tools. This demonstrates SimilarGPT's practicality and efficiency in smart contract security analysis.
    \end{mdframed}

\subsection{RQ2 - Socratic Debate Framework}

The two primary components of the SimilarGPT are vulnerability identification and false positive filtering. While part ot CBSC identifies vulnerabilities, the \textit{Socratic method} is responsible for mitigating the impact of the LLM's hallucination. This section uses the \textit{Solodit} dataset that was previously mentioned in \ref{subsection:setup}.

The \textit{Socratic method}\cite{soc_solo,soc_average} is a type of debate method that involves asking and answering questions, beginning with universal beliefs and examining their internal coherence through rebuttals. Here, we attempt to demonstrate the \textit{Socratic method}'s effectiveness in filtering false positives. We refer to the study \cite{Do_you_still_need,gptlens} as a comparison. Specifically, we construct two frameworks. 1) a conventional one-stage framework; 2) a two-stage framework. In the one-stage framework\cite{Do_you_still_need}, both vulnerability detection and interpretation are performed by the same agent; however, in the latter two-stage framework\cite{gptlens,gptscan}, former agent is responsible for detecting the vulnerability, while the latter is responsible for determining the validity of the vulnerability interpretation. The prompts in two of these frameworks are similar to SimilarGPT, and the same code similarity-based detection mechanism is employed. 

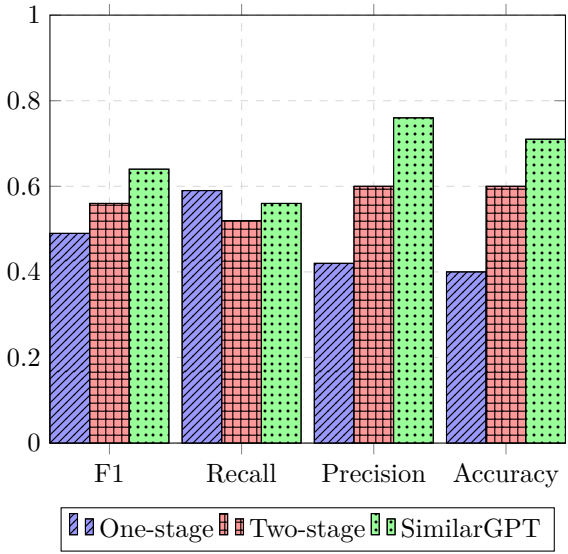
\begin{figure}[ht]
    \centering
    \begin{tikzpicture}
    \begin{axis}[
        ybar=0pt, 
        bar width=15pt, 
        enlarge x limits=0.15, 
        symbolic x coords={F1, Recall, Precision, Accuracy}, 
        xtick=data,
        ymin=0, ymax=1, 
        legend style={at={(0.5,-0.15)}, 
        anchor=north,legend columns=-1},
        nodes near coords align={vertical}, 
        every node near coord/.append style={font=\tiny}, 
        yticklabel style={/pgf/number format/fixed, /pgf/number format/precision=2}, 
        grid=major, 
        grid style={dashed,gray!30} 
    ]
    \addplot[fill=blue!40, postaction={pattern=north east lines}] coordinates {(F1,0.49) (Recall,0.59) (Precision,0.42) (Accuracy,0.40)}; 
    \addplot[fill=red!40, postaction={pattern=grid}] coordinates {(F1,0.56) (Recall,0.52) (Precision,0.60) (Accuracy,0.60)}; 
    \addplot[fill=green!40, postaction={pattern=dots}] coordinates {(F1,0.64) (Recall,0.56) (Precision,0.76) (Accuracy,0.71)}; 
    \legend{One-stage, Two-stage, SimilarGPT} 
    \end{axis}
    \end{tikzpicture}
    \captionsetup{justification=centering} 
    \caption{Comparing SimilarGPT with the traditional integration models.} 
    \label{fig:sug1}    
\end{figure}

In the results shown in Fig. \ref{fig:sug1}, our approach outperforms the traditional model in all four main performance metrics. The fact is that the three frameworks do not differ much in recall rate. Even the single-agent framework performs better on recall. However, when further analyzed, we find that their frameworks do not perform well on negative samples, resulting in the highest false positive rate, reaching 57\% false positive rate, for the one-stage. Observing the experimental results suggests this issue is primarily due to hallucinations in LLMs.
\begin{mdframed}[backgroundcolor=gray!30, hidealllines=true] 
    \textbf{Answer for RQ2:}the application of the Socratic Debate Framework in SimilarGPT reduced the false positive rate to 12\%, compared to 57\% in single-agent frameworks. This demonstrates the framework's effectiveness in refining the detection process by mitigating the influence of hallucinations in large language models.
    \end{mdframed}

\subsection{RQ3 - Code-based Similarity Checking}
In this section, we try to answer question for \textit{how effective CBSC is in improving the recall rate}.

The dataset for the experiment is the \textit{Solodit dataset}, which contains 138 samples, of which 67 are positive examples and 71 are negative examples. This dataset is also used in the ablation experiment of the Socratic method.

We have implemented modifications to the framework. In the detector section, we have eliminated the CBSC  method and instead used the code to be detected as input to the detector. Additionally, the default settings are maintained for the model parameters, except for the temperature option.

\begin{table}[ht]
    \centering
    \begin{tabular}{c | c | c | c | c | c} 
     \hline
      & TP  & TN  & FP  & FN  & Sum \\ [0.5ex] 
     \hline
    with SimilarChecking & 38 & 63  & 12 & 30 & 143 \\ 
    without SimilarChecking & 20 & 61  & 10 & 47 & 138 \\ 
    \hline
    \end{tabular}
    \caption{detecting results before and after SimilarChecking.}
    \label{table:similar_checking}
    \end{table}

    Table \ref{table:similar_checking} shows that in the \textit{Solodit dataset}, SimilarGPT discovered 38 true positives (TPs) and produced 12 false positives (FPs). Without a CBSC method, there are only 20 TPs, which is roughly half of the former. However, we highlight that in the absence of CBSC method. The precision rate is 66\%. There isn't much of a difference between these examples and the ones with CBSC method. A plausible assumption is that CBSC method has no substantial influence on precision rates.

\begin{mdframed}[backgroundcolor=gray!30, hidealllines=true]
    \textbf{Answer for RQ3:}the use of Code-based similarity checking in SimilarGPT raised the true positive detections from 20 to 38, highlighting its effectiveness in improving the recall rate and thereby enhancing the tool's overall accuracy in identifying vulnerabilities.
    \end{mdframed}

\section{RELATED WORK}
\label{sec:related_work}
\textbf{Vulnerability Detection}. In recent years, vulnerability detection has received a lot of attention, especially given the rapid development of blockchain and smart contract technologies. Early research concentrated on approaches like static analysis, dynamic analysis, and formal verification. And for those static analysis tools, such as \textit{Slither} \cite{slither}, Oyente \cite{oyente}, and others \cite{brent2020ethainter, brent2018vandal,kalra2018zeus}, identify possible vulnerabilities by comparing the code's syntax and control flow to preset rules. However, such approaches have limitations in dealing with complex logic, dynamic behavior, and low-level vulnerabilities (e.g., re-entry attacks), making it difficult to detect complex logic errors and resulting in too many false positives. Dynamic analysis, like Fuzzy testing Confuzzius\cite{Confuzzius_paper}, Sfuzz\cite{sfuzz}, and other tools\cite{jiang2018contractfuzzer} by automatically generating inputs to test the behavior of the system, as well as symbolic execution Manticore\ cite{manticore}, \textit{Mythril}\cite{mythril},\cite{liu2022finding,wang2020oracle,frank2020ethbmc}. Formal verification \cite{permenev2020verx,so2020verismart} ensures the correctness of the code through mathematical proofs. While providing a high degree of accuracy, it is usually applicable to smaller or relatively simple systems, and the verification process is complex and time-consuming. 

Based on the observation that the code reuse rate is high on Ethernet \cite{code_reuse,111,112}, vulnerability detection methods based on code similarity have been properly developed. Liu\cite{liu2018eclone} et al, developed a semantic clone detection tool for Smart Contracts in Ethereum. The method captures essential semantic elements from symbolic transaction execution and converts them into vectors for similarity evaluations. To address the constraints of vulnerability detection, they offer a semantic-aware security auditing approach \cite{liu2018s} that evaluates vulnerabilities through N-gram language modeling and static semantic labeling. In addition, \cite{liu2019enabling} enhanced vulnerability detection accuracy by depicting syntactic and semantic aspects of contracts using the birthmark. Huang\cite{huang2021hunting} et al, analyzed data flow and control flow using key instructions and the Graph2Vector tool for similarity. Pierro\cite{pierro2021analysis}, Chen\cite{chen2021understanding}, and Gao\cite{gao2020checking} evaluated contract similarity and detected vulnerabilities by analyzing the edit distance of Abstract Syntax Trees (AST), node type hash sequences, and normalization of variables and constant values, respectively. These studies collectively advanced the development of smart contract security analysis.

\text{LLM-based vulnerability detection}.Researchers have made extensive use of Large Language Models (LLMs) for vulnerability detection, including Ullah\cite{ullah2023llms}, Fu\cite{fu2023chatgpt}, Thapa\cite{thapa2022transformer}, David\cite{ david2023you}, Alqarni\cite{alqarni2022low}, Sun\cite{sun2024llm4vuln}, Mathews\cite{mathews2024llbezpeky}, Hu\cite{hu2023large} and Purba \cite{purba2023software}, among others. They evaluated the performance of LLMs on vulnerability detection tasks, analyzed the gaps, and proposed ways to improve detection capabilities. Sun\cite{sun2024gptscan} et al. proposed GPTScan, which combines LLMs with static program analysis, while Li\cite{li2023hitchhiker} et al. proposed LLift, which integrates LLMs with static analysis tools to improve the accuracy of detecting logic vulnerabilities. In addition, LLMs have been used for other security tasks, such as TitanFuzz\cite{deng2023large} and FuzzGPT\cite{deng2024large} by Deng et al. In addition to ChatAFL\cite{meng2024large} for fuzzing and protocol testing. LLMs are also applied to program repair tasks such as ACFix\cite{zhang2024acfix} and ChatRepair.

\section{THREATS TO VALIDITY}
\label{sec:THREATS}
Our study currently may have two limitations. The first involves the randomness of the LLM's output. To reduce the bias produced by different parameters and prompts, we try to use default settings for models. We also analyze model performance using several runs and average the findings to assure the stability and reproducibility of the outcomes. The second constraint is the danger of data bias during the data gathering procedure, which may have an impact on the model evaluation outcomes. To address this issue, we employed a variety of data sources during the data gathering phase and rigorously screened and cleaned the data to reduce the influence of bias. In addition, we intend to include more control variables and experimental designs in future studies to help validate our findings. Nonetheless, these limitations remind us that we must exercise caution when interpreting our findings and conduct more extensive validation and cross-validation where possible to ensure the reliability and generalizability of our findings. We expect to address these constraints by continuously optimizing our methodology and tools in future works.

\section{CONCLUSION}
\label{sec:conclusion}
In this research, we introduce SimilarGPT, a smart contract vulnerability detection tool that integrates Large Language Models with Code-based similarity checking. SimilarGPT effectively identify vulnerabilities in smart contracts by leveraging Ethereum's prevalent code reuse issue. Our experimental results reveal that SimilarGPT excels at increasing the recall rate of vulnerability detection while decreasing the false positive rate. In particular, by implementing the Socratic method, we significantly mitigating the false positives caused by LLMs hallucination. Furthermore, SimilarGPT allows it to adapt to the changing security ecosystem of smart contracts. Future work will focus on enhancing the code similarity identification method and expanding the dataset to increase the tool's detection accuracy and usefulness.


\bibliographystyle{IEEEtran}
\bibliography{Mybib}

\begin{thebibliography}{10}
\providecommand{\url}[1]{#1}
\csname url@samestyle\endcsname
\providecommand{\newblock}{\relax}
\providecommand{\bibinfo}[2]{#2}
\providecommand{\BIBentrySTDinterwordspacing}{\spaceskip=0pt\relax}
\providecommand{\BIBentryALTinterwordstretchfactor}{4}
\providecommand{\BIBentryALTinterwordspacing}{\spaceskip=\fontdimen2\font plus
\BIBentryALTinterwordstretchfactor\fontdimen3\font minus \fontdimen4\font\relax}
\providecommand{\BIBforeignlanguage}[2]{{%
\expandafter\ifx\csname l@#1\endcsname\relax
\typeout{** WARNING: IEEEtran.bst: No hyphenation pattern has been}%
\typeout{** loaded for the language `#1'. Using the pattern for}%
\typeout{** the default language instead.}%
\else
\language=\csname l@#1\endcsname
\fi
#2}}
\providecommand{\BIBdecl}{\relax}
\BIBdecl

\bibitem{ethereum_main_page}
\url{https://ethereum.org/zh/}, accessed November 4, 2024.

\bibitem{defillama_totallocked}
\url{https://defillama.com/}, accessed November 4, 2024.

\bibitem{defillama_hacks}
\url{https://defillama.com/hacks}, accessed November 4, 2024.

\bibitem{web3bugs}
Z.~Zhang, B.~Zhang, W.~Xu, and Z.~Lin, ``Demystifying exploitable bugs in smart contracts,'' in \emph{2023 IEEE/ACM 45th International Conference on Software Engineering (ICSE)}.\hskip 1em plus 0.5em minus 0.4em\relax IEEE, 2023, pp. 615--627.

\bibitem{mythril}
\url{https://github.com/Consensys/mythril}, accessed November 4, 2024.

\bibitem{slither}
\url{https://github.com/crytic/slither}, accessed November 4, 2024.

\bibitem{Confuzzius_paper}
C.~F. Torres, A.~K. Iannillo, A.~Gervais, and R.~State, ``Confuzzius: A data dependency-aware hybrid fuzzer for smart contracts,'' in \emph{2021 IEEE European Symposium on Security and Privacy (EuroS\&P)}.\hskip 1em plus 0.5em minus 0.4em\relax IEEE, 2021, pp. 103--119.

\bibitem{oyente}
\url{https://github.com/enzymefinance/oyente}, accessed November 4, 2024.

\bibitem{rodler2018sereum}
M.~Rodler, W.~Li, G.~O. Karame, and L.~Davi, ``Sereum: Protecting existing smart contracts against re-entrancy attacks,'' \emph{arXiv preprint arXiv:1812.05934}, 2018.

\bibitem{tan2022soltype}
B.~Tan, B.~Mariano, S.~K. Lahiri, I.~Dillig, and Y.~Feng, ``Soltype: refinement types for arithmetic overflow in solidity,'' \emph{Proceedings of the ACM on Programming Languages}, vol.~6, no. POPL, pp. 1--29, 2022.

\bibitem{chaliasos2024smart}
S.~Chaliasos, M.~A. Charalambous, L.~Zhou, R.~Galanopoulou, A.~Gervais, D.~Mitropoulos, and B.~Livshits, ``Smart contract and defi security tools: Do they meet the needs of practitioners?'' in \emph{Proceedings of the 46th IEEE/ACM International Conference on Software Engineering}, 2024, pp. 1--13.

\bibitem{gpt-4_report}
J.~Achiam, S.~Adler, S.~Agarwal, L.~Ahmad, I.~Akkaya, F.~L. Aleman, D.~Almeida, J.~Altenschmidt, S.~Altman, S.~Anadkat \emph{et~al.}, ``Gpt-4 technical report,'' \emph{arXiv preprint arXiv:2303.08774}, 2023.

\bibitem{A_survey_of_large_language_models}
W.~X. Zhao, K.~Zhou, J.~Li, T.~Tang, X.~Wang, Y.~Hou, Y.~Min, B.~Zhang, J.~Zhang, Z.~Dong \emph{et~al.}, ``A survey of large language models,'' \emph{arXiv preprint arXiv:2303.18223}, 2023.

\bibitem{gptscan}
Y.~Sun, D.~Wu, Y.~Xue, H.~Liu, H.~Wang, Z.~Xu, X.~Xie, and Y.~Liu, ``Gptscan: Detecting logic vulnerabilities in smart contracts by combining gpt with program analysis,'' in \emph{Proceedings of the IEEE/ACM 46th International Conference on Software Engineering}, 2024, pp. 1--13.

\bibitem{gptlens}
S.~Hu, T.~Huang, F.~{\.I}lhan, S.~F. Tekin, and L.~Liu, ``Large language model-powered smart contract vulnerability detection: New perspectives,'' in \emph{2023 5th IEEE International Conference on Trust, Privacy and Security in Intelligent Systems and Applications (TPS-ISA)}.\hskip 1em plus 0.5em minus 0.4em\relax IEEE, 2023, pp. 297--306.

\bibitem{llm4vuln}
Y.~Sun, D.~Wu, Y.~Xue, H.~Liu, W.~Ma, L.~Zhang, Y.~Liu, and Y.~Li, ``Llm4vuln: A unified evaluation framework for decoupling and enhancing llms' vulnerability reasoning,'' \emph{arXiv preprint arXiv:2401.16185}, 2024.

\bibitem{propertygpt}
Y.~Liu, Y.~Xue, D.~Wu, Y.~Sun, Y.~Li, M.~Shi, and Y.~Liu, ``Propertygpt: Llm-driven formal verification of smart contracts through retrieval-augmented property generation,'' \emph{arXiv preprint arXiv:2405.02580}, 2024.

\bibitem{Do_you_still_need}
I.~David, L.~Zhou, K.~Qin, D.~Song, L.~Cavallaro, and A.~Gervais, ``Do you still need a manual smart contract audit?'' \emph{arXiv preprint arXiv:2306.12338}, 2023.

\bibitem{111}
F.~Khan, I.~David, D.~Varro, and S.~McIntosh, ``Code cloning in smart contracts on the ethereum platform: An extended replication study,'' \emph{IEEE Transactions on Software Engineering}, vol.~49, no.~4, pp. 2006--2019, 2022.

\bibitem{112}
M.~Kondo, G.~A. Oliva, Z.~M. Jiang, A.~E. Hassan, and O.~Mizuno, ``Code cloning in smart contracts: a case study on verified contracts from the ethereum blockchain platform,'' \emph{Empirical Software Engineering}, vol.~25, pp. 4617--4675, 2020.

\bibitem{code_reuse}
K.~Sun, Z.~Xu, C.~Liu, K.~Li, and Y.~Liu, ``Demystifying the composition and code reuse in solidity smart contracts,'' in \emph{Proceedings of the 31st ACM Joint European Software Engineering Conference and Symposium on the Foundations of Software Engineering}, 2023, pp. 796--807.

\bibitem{huang2021hunting}
J.~Huang, S.~Han, W.~You, W.~Shi, B.~Liang, J.~Wu, and Y.~Wu, ``Hunting vulnerable smart contracts via graph embedding based bytecode matching,'' \emph{IEEE Transactions on Information Forensics and Security}, vol.~16, pp. 2144--2156, 2021.

\bibitem{liu2018eclone}
H.~Liu, Z.~Yang, C.~Liu, Y.~Jiang, W.~Zhao, and J.~Sun, ``Eclone: Detect semantic clones in ethereum via symbolic transaction sketch,'' in \emph{Proceedings of the 2018 26th ACM Joint Meeting on European Software Engineering Conference and Symposium on the Foundations of Software Engineering}, 2018, pp. 900--903.

\bibitem{smartembed}
Z.~Gao, L.~Jiang, X.~Xia, D.~Lo, and J.~Grundy, ``Checking smart contracts with structural code embedding,'' \emph{IEEE Transactions on Software Engineering}, vol.~47, no.~12, pp. 2874--2891, 2020.

\bibitem{defi}
F.~Sch{\"a}r, ``Decentralized finance: On blockchain-and smart contract-based financial markets,'' \emph{FRB of St. Louis Review}, 2021.

\bibitem{zhou2023sok}
L.~Zhou, X.~Xiong, J.~Ernstberger, S.~Chaliasos, Z.~Wang, Y.~Wang, K.~Qin, R.~Wattenhofer, D.~Song, and A.~Gervais, ``Sok: Decentralized finance (defi) attacks,'' in \emph{2023 IEEE Symposium on Security and Privacy (SP)}.\hskip 1em plus 0.5em minus 0.4em\relax IEEE, 2023, pp. 2444--2461.

\bibitem{zero_shot}
T.~Kojima, S.~S. Gu, M.~Reid, Y.~Matsuo, and Y.~Iwasawa, ``Large language models are zero-shot reasoners,'' \emph{Advances in neural information processing systems}, vol.~35, pp. 22\,199--22\,213, 2022.

\bibitem{sun2024llm4vuln}
Y.~Sun, D.~Wu, Y.~Xue, H.~Liu, W.~Ma, L.~Zhang, Y.~Liu, and Y.~Li, ``Llm4vuln: A unified evaluation framework for decoupling and enhancing llms' vulnerability reasoning,'' \emph{arXiv preprint arXiv:2401.16185}, 2024.

\bibitem{_hyperparameters}
Z.~Lin, M.~Feng, C.~N.~d. Santos, M.~Yu, B.~Xiang, B.~Zhou, and Y.~Bengio, ``A structured self-attentive sentence embedding,'' \emph{arXiv preprint arXiv:1703.03130}, 2017.

\bibitem{_rag}
P.~Lewis, E.~Perez, A.~Piktus, F.~Petroni, V.~Karpukhin, N.~Goyal, H.~K{\"u}ttler, M.~Lewis, W.-t. Yih, T.~Rockt{\"a}schel \emph{et~al.}, ``Retrieval-augmented generation for knowledge-intensive nlp tasks,'' \emph{Advances in Neural Information Processing Systems}, vol.~33, pp. 9459--9474, 2020.

\bibitem{llm_hallucination}
Z.~Ji, N.~Lee, R.~Frieske, T.~Yu, D.~Su, Y.~Xu, E.~Ishii, Y.~J. Bang, A.~Madotto, and P.~Fung, ``Survey of hallucination in natural language generation,'' \emph{ACM Computing Surveys}, vol.~55, no.~12, pp. 1--38, 2023.

\bibitem{immunefi}
\url{https://immunefi.com/}, accessed November 4, 2024.

\bibitem{topo}
\url{https://www.wikiwand.com/en/articles/Topological_sorting}, accessed November 4, 2024.

\bibitem{surya}
\url{https://github.com/ConsenSys/surya}, accessed November 4, 2024.

\bibitem{openzeppelin}
\url{https://www.npmjs.com/package/@openzeppelin/contracts}, accessed November 4, 2024.

\bibitem{huggingface_all-MiniLM-L6-v2}
\url{https://huggingface.co/sentence-transformers/all-MiniLM-L6-v2}, accessed November 4, 2024.

\bibitem{sentence_transformer}
\url{https://www.sbert.net/docs/sentence_transformer/pretrained_models.html#scientific-similarity-models}, accessed November 4, 2024.

\bibitem{sbert_all-MiniLM-L6-v2}
\url{https://www.sbert.net/docs/sentence_transformer/pretrained_models.html}, accessed November 4, 2024.

\bibitem{sbert_1}
\url{https://huggingface.co/sentence-transformers/all-mpnet-base-v2}, accessed November 4, 2024.

\bibitem{sbert_2}
\url{https://huggingface.co/sentence-transformers/multi-qa-mpnet-base-dot-v1}, accessed November 4, 2024.

\bibitem{sbert_3}
\url{https://huggingface.co/sentence-transformers/multi-qa-MiniLM-L6-cos-v1}, accessed November 4, 2024.

\bibitem{defihack}
\url{https://github.com/SunWeb3Sec/DeFiHackLabs?tab=readme-ov-file}, accessed November 4, 2024.

\bibitem{slowmist}
\url{https://hacked.slowmist.io/en/}, accessed November 4, 2024.

\bibitem{github}
\url{https://github.com/}, accessed November 4, 2024.

\bibitem{soc_average}
J.~Qi, Z.~Xu, Y.~Shen, M.~Liu, D.~Jin, Q.~Wang, and L.~Huang, ``The art of socratic questioning: Recursive thinking with large language models,'' \emph{arXiv preprint arXiv:2305.14999}, 2023.

\bibitem{soc_Markov}
X.~Sun, J.~Li, Y.~Zhong, D.~Zhao, and R.~Yan, ``Towards detecting llms hallucination via markov chain-based multi-agent debate framework,'' \emph{arXiv preprint arXiv:2406.03075}, 2024.

\bibitem{soc_solo}
E.~Y. Chang, ``Prompting large language models with the socratic method,'' in \emph{2023 IEEE 13th Annual Computing and Communication Workshop and Conference (CCWC)}.\hskip 1em plus 0.5em minus 0.4em\relax IEEE, 2023, pp. 0351--0360.

\bibitem{soc_others}
T.~Chowdhury, C.~Ling, X.~Zhang, X.~Zhao, G.~Bai, J.~Pei, H.~Chen, and L.~Zhao, ``Knowledge-enhanced neural machine reasoning: A review,'' \emph{arXiv preprint arXiv:2302.02093}, 2023.

\bibitem{cve}
\url{https://cve.mitre.org/cgi-bin/cvekey.cgi?keyword=smart+contract}, accessed November 4, 2024.

\bibitem{solodit}
\url{https://solodit.xyz/}, accessed November 4, 2024.

\bibitem{uranium}
\url{https://github.com/SunWeb3Sec/DeFiHackLabs/blob/main/past/2021/README.md#20210428-uranium---miscalculation}, accessed November 4, 2024.

\bibitem{Nimbus}
\url{https://github.com/SunWeb3Sec/DeFiHackLabs/blob/main/past/2021/README.md#20210915-nimbus-platform}, accessed November 4, 2024.

\bibitem{swapos}
\url{https://github.com/SunWeb3Sec/DeFiHackLabs/blob/main/past/2023/README.md#20230416-swapos-v2---error-k-value-attack}, accessed November 4, 2024.

\bibitem{brent2020ethainter}
L.~Brent, N.~Grech, S.~Lagouvardos, B.~Scholz, and Y.~Smaragdakis, ``Ethainter: a smart contract security analyzer for composite vulnerabilities,'' in \emph{Proceedings of the 41st ACM SIGPLAN Conference on Programming Language Design and Implementation}, 2020, pp. 454--469.

\bibitem{brent2018vandal}
L.~Brent, A.~Jurisevic, M.~Kong, E.~Liu, F.~Gauthier, V.~Gramoli, R.~Holz, and B.~Scholz, ``Vandal: A scalable security analysis framework for smart contracts,'' \emph{arXiv preprint arXiv:1809.03981}, 2018.

\bibitem{kalra2018zeus}
S.~Kalra, S.~Goel, M.~Dhawan, and S.~Sharma, ``Zeus: analyzing safety of smart contracts.'' in \emph{Ndss}, 2018, pp. 1--12.

\bibitem{sfuzz}
T.~D. Nguyen, L.~H. Pham, J.~Sun, Y.~Lin, and Q.~T. Minh, ``sfuzz: An efficient adaptive fuzzer for solidity smart contracts,'' in \emph{Proceedings of the ACM/IEEE 42nd International Conference on Software Engineering}, 2020, pp. 778--788.

\bibitem{jiang2018contractfuzzer}
B.~Jiang, Y.~Liu, and W.~K. Chan, ``Contractfuzzer: Fuzzing smart contracts for vulnerability detection,'' in \emph{Proceedings of the 33rd ACM/IEEE international conference on automated software engineering}, 2018, pp. 259--269.

\bibitem{liu2022finding}
Y.~Liu, Y.~Li, S.-W. Lin, and C.~Artho, ``Finding permission bugs in smart contracts with role mining,'' in \emph{Proceedings of the 31st ACM SIGSOFT International Symposium on Software Testing and Analysis}, 2022, pp. 716--727.

\bibitem{wang2020oracle}
H.~Wang, Y.~Liu, Y.~Li, S.-W. Lin, C.~Artho, L.~Ma, and Y.~Liu, ``Oracle-supported dynamic exploit generation for smart contracts,'' \emph{IEEE Transactions on Dependable and Secure Computing}, vol.~19, no.~3, pp. 1795--1809, 2020.

\bibitem{frank2020ethbmc}
J.~Frank, C.~Aschermann, and T.~Holz, ``$\{$ETHBMC$\}$: A bounded model checker for smart contracts,'' in \emph{29th USENIX Security Symposium (USENIX Security 20)}, 2020, pp. 2757--2774.

\bibitem{permenev2020verx}
A.~Permenev, D.~Dimitrov, P.~Tsankov, D.~Drachsler-Cohen, and M.~Vechev, ``Verx: Safety verification of smart contracts,'' in \emph{2020 IEEE symposium on security and privacy (SP)}.\hskip 1em plus 0.5em minus 0.4em\relax IEEE, 2020, pp. 1661--1677.

\bibitem{so2020verismart}
S.~So, M.~Lee, J.~Park, H.~Lee, and H.~Oh, ``Verismart: A highly precise safety verifier for ethereum smart contracts,'' in \emph{2020 IEEE Symposium on Security and Privacy (SP)}.\hskip 1em plus 0.5em minus 0.4em\relax IEEE, 2020, pp. 1678--1694.

\bibitem{liu2018s}
H.~Liu, C.~Liu, W.~Zhao, Y.~Jiang, and J.~Sun, ``S-gram: towards semantic-aware security auditing for ethereum smart contracts,'' in \emph{Proceedings of the 33rd ACM/IEEE international conference on automated software engineering}, 2018, pp. 814--819.

\bibitem{liu2019enabling}
H.~Liu, Z.~Yang, Y.~Jiang, W.~Zhao, and J.~Sun, ``Enabling clone detection for ethereum via smart contract birthmarks,'' in \emph{2019 IEEE/ACM 27th International Conference on Program Comprehension (ICPC)}.\hskip 1em plus 0.5em minus 0.4em\relax IEEE, 2019, pp. 105--115.

\bibitem{pierro2021analysis}
G.~A. Pierro and R.~Tonelli, ``Analysis of source code duplication in ethreum smart contracts,'' in \emph{2021 IEEE International Conference on Software Analysis, Evolution and Reengineering (SANER)}.\hskip 1em plus 0.5em minus 0.4em\relax IEEE, 2021, pp. 701--707.

\bibitem{chen2021understanding}
X.~Chen, P.~Liao, Y.~Zhang, Y.~Huang, and Z.~Zheng, ``Understanding code reuse in smart contracts,'' in \emph{2021 IEEE International Conference on Software Analysis, Evolution and Reengineering (SANER)}.\hskip 1em plus 0.5em minus 0.4em\relax IEEE, 2021, pp. 470--479.

\bibitem{gao2020checking}
Z.~Gao, L.~Jiang, X.~Xia, D.~Lo, and J.~Grundy, ``Checking smart contracts with structural code embedding,'' \emph{IEEE Transactions on Software Engineering}, vol.~47, no.~12, pp. 2874--2891, 2020.

\bibitem{ullah2023llms}
S.~Ullah, M.~Han, S.~Pujar, H.~Pearce, A.~Coskun, and G.~Stringhini, ``Llms cannot reliably identify and reason about security vulnerabilities (yet?): A comprehensive evaluation, framework, and benchmarks,'' \emph{arXiv preprint arXiv:2312.12575}, 2023.

\bibitem{fu2023chatgpt}
M.~Fu, C.~K. Tantithamthavorn, V.~Nguyen, and T.~Le, ``Chatgpt for vulnerability detection, classification, and repair: How far are we?'' in \emph{2023 30th Asia-Pacific Software Engineering Conference (APSEC)}.\hskip 1em plus 0.5em minus 0.4em\relax IEEE, 2023, pp. 632--636.

\bibitem{thapa2022transformer}
C.~Thapa, S.~I. Jang, M.~E. Ahmed, S.~Camtepe, J.~Pieprzyk, and S.~Nepal, ``Transformer-based language models for software vulnerability detection,'' in \emph{Proceedings of the 38th Annual Computer Security Applications Conference}, 2022, pp. 481--496.

\bibitem{david2023you}
I.~David, L.~Zhou, K.~Qin, D.~Song, L.~Cavallaro, and A.~Gervais, ``Do you still need a manual smart contract audit?'' \emph{arXiv preprint arXiv:2306.12338}, 2023.

\bibitem{alqarni2022low}
M.~Alqarni and A.~Azim, ``Low level source code vulnerability detection using advanced bert language model.'' in \emph{Canadian AI}, 2022.

\bibitem{mathews2024llbezpeky}
N.~S. Mathews, Y.~Brus, Y.~Aafer, M.~Nagappan, and S.~McIntosh, ``Llbezpeky: Leveraging large language models for vulnerability detection,'' \emph{arXiv preprint arXiv:2401.01269}, 2024.

\bibitem{hu2023large}
S.~Hu, T.~Huang, F.~{\.I}lhan, S.~F. Tekin, and L.~Liu, ``Large language model-powered smart contract vulnerability detection: New perspectives,'' in \emph{2023 5th IEEE International Conference on Trust, Privacy and Security in Intelligent Systems and Applications (TPS-ISA)}.\hskip 1em plus 0.5em minus 0.4em\relax IEEE, 2023, pp. 297--306.

\bibitem{purba2023software}
M.~D. Purba, A.~Ghosh, B.~J. Radford, and B.~Chu, ``Software vulnerability detection using large language models,'' in \emph{2023 IEEE 34th International Symposium on Software Reliability Engineering Workshops (ISSREW)}.\hskip 1em plus 0.5em minus 0.4em\relax IEEE, 2023, pp. 112--119.

\bibitem{sun2024gptscan}
Y.~Sun, D.~Wu, Y.~Xue, H.~Liu, H.~Wang, Z.~Xu, X.~Xie, and Y.~Liu, ``Gptscan: Detecting logic vulnerabilities in smart contracts by combining gpt with program analysis,'' in \emph{Proceedings of the IEEE/ACM 46th International Conference on Software Engineering}, 2024, pp. 1--13.

\bibitem{li2023hitchhiker}
H.~Li, Y.~Hao, Y.~Zhai, and Z.~Qian, ``The hitchhiker's guide to program analysis: A journey with large language models,'' \emph{arXiv preprint arXiv:2308.00245}, 2023.

\bibitem{deng2023large}
Y.~Deng, C.~S. Xia, H.~Peng, C.~Yang, and L.~Zhang, ``Large language models are zero-shot fuzzers: Fuzzing deep-learning libraries via large language models,'' in \emph{Proceedings of the 32nd ACM SIGSOFT international symposium on software testing and analysis}, 2023, pp. 423--435.

\bibitem{deng2024large}
Y.~Deng, C.~S. Xia, C.~Yang, S.~D. Zhang, S.~Yang, and L.~Zhang, ``Large language models are edge-case generators: Crafting unusual programs for fuzzing deep learning libraries,'' in \emph{Proceedings of the 46th IEEE/ACM International Conference on Software Engineering}, 2024, pp. 1--13.

\bibitem{meng2024large}
R.~Meng, M.~Mirchev, M.~B{\"o}hme, and A.~Roychoudhury, ``Large language model guided protocol fuzzing,'' in \emph{Proceedings of the 31st Annual Network and Distributed System Security Symposium (NDSS)}, 2024.

\bibitem{zhang2024acfix}
L.~Zhang, K.~Li, K.~Sun, D.~Wu, Y.~Liu, H.~Tian, and Y.~Liu, ``Acfix: Guiding llms with mined common rbac practices for context-aware repair of access control vulnerabilities in smart contracts,'' \emph{arXiv preprint arXiv:2403.06838}, 2024.

\end{thebibliography}
\vspace{12pt}
\color{red}

\end{document}